\documentclass[aps,prb,twocolumn,showpacs,10pt,superscriptaddress]{revtex4-1}

\usepackage{graphicx}
\usepackage{subfigure}
\usepackage{amsmath}
\usepackage{amsfonts}
\usepackage{amssymb}
\usepackage{mathrsfs}
\usepackage{bbm}
\usepackage{subfigure}
\usepackage{xcolor}

\begin{document}

\title{Proposal for a flux qubit in a dc SQUID with the $4\pi$ period Josephson effect}
\author{Wen-Chao Huang}
\affiliation{School of Physics and Engineering, Sun Yat-sen University, Guangzhou 510275, China}

\author{Qi-Feng Liang}
\affiliation{Department of Physics, Shaoxing University, Shaoxing 312000, China}

\author{Dao-Xin Yao}
\email{yaodaox@mail.sysu.edu.cn}
\affiliation{School of Physics and Engineering, Sun Yat-sen University, Guangzhou 510275, China}

\author{Zhi Wang}
\email{wangzh356@mail.sysu.edu.cn}
\affiliation{School of Physics and Engineering, Sun Yat-sen University, Guangzhou 510275, China}

\begin{abstract}
Constructing qubits which are suitable for quantum computation remains a notable challenge. Here, we propose a superconducting flux qubit in a dc SQUID structure, formed by a conventional insulator Josephson junction and a topological nanowire Josephson junction with Majorana bound states. The zero energy Majorana bound states transport $4\pi$ period Josephson currents in the nanowire junction. The interplay between this $4\pi$ period Josephson effect and the convectional $2\pi$ period Josephson effect in the insulator junction induces a double-well potential energy landscape in the SQUID. As a result, the two lowest energy levels of the SQUID are isolated from other levels.
These two levels show contradicting circulating supercurrents, thus can be used as a flux qubit. We reveal that this flux qubit has the merits of stability to external noises, tolerance to the deviation of system parameters, and scalability to large numbers. Furthermore, we demonstrate how to couple this flux qubit with the Majorana qubit by tuning the junction parameters, and how to use this coupling to manipulate the Majorana qubit.
\end{abstract}

\maketitle


\section{Introduction}
Quantum computation, which involves processing information on quantum variables, is valuable in directly simulating quantum mechanics or efficiently solving classical problems \cite{Nielsen,DiVincenzo}.
The realization of the quantum computation calls for the coherent storage and manipulation of the quantum information\cite{Lloyd}.
A typical scheme for coherent quantum information processing consists of two energy level systems, the so-called quantum bits (qubits)\cite{Schumacher}. In a qubit,
the two levels serve as the basis states, then the quantum superposition of these basis states can store the quantum information.
A number of qubits have been proposed in both natural and artificial two-level systems\cite{spin,QED,charge,flux3,flux1}.
In particular, the artificial qubits seem to be very promising for realistic application. In fact, more than one hundred artificial qubits have been integrated in the recent commercial quantum computers\cite{dwave}.

\begin{figure}[b]
\begin{center}
\includegraphics[clip = true, width = \columnwidth]{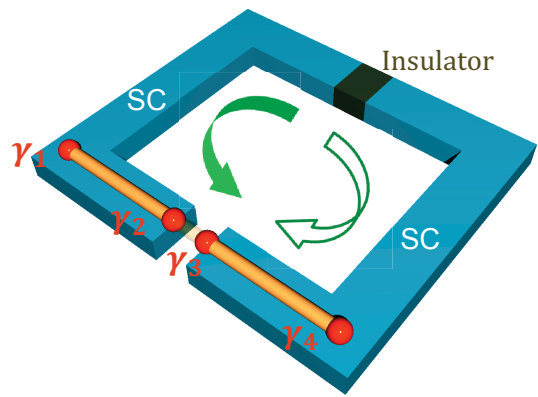}
\caption{(Color online). Schematic setup of a flux qubit in a dc SQUID, formed by a conventional insulator Josephson junction and a topological nanowire Josephson junction. The wire hosts four Majorana bound states $\gamma_{1,2,3,4}$. The two arrows indicate the two contradicting circulating directions of the supercurrents, which are the basis states of the qubit. }
\end{center}
\end{figure}

One attractive candidate among the artificial qubits is the superconducting flux qubit\cite{fluxrmp,nori,clarke,devoret}, which is composed of a superconducting loop interrupted by one or more Josephson junctions.
Flux qubits have several outstanding advantages\cite{fluxrmp,nori}. The first one is the scalability: flux qubits are based on lithography technology which allows a quantum computer to integrate a large number of qubits. Second, flux qubits are resistent to
the setback charge fluctuation, which is the major noise in the electronic systems.
The last and most important advantage is the flexibility: the property of the the flux qubit is determined by the design of the Josephson junction system, in which the parameters can be tuned continuously.
Two designs for flux qubits have been studied up to now\cite{flux1,flux3}: (i) an rf SQUID with one Josephson junction; and (ii) a superconducting ring with three Josephson junctions.
The one-junction design demands a high self-inductance of $L \sim 200p{\rm H}$, which can be realized only in a large device. For this reason, the qubit is usually very susceptible to external noises\cite{fluxrmp}.
The three-junction qubit is constructed in a superconducting loop of micrometer size with a very small self-inductance $L \sim 5p{\rm H}$ and low critical currents $I_c \sim 200 {\rm nA}$. Therefore, this qubit has weak coupling with external fluctuation fields and was adopted more widely\cite{clarke}.
In this three-junction design, however, the parameters of the junctions are severely confined. Two of the three junctions must have exactly the same critical current while the third junction has a slightly smaller critical current. This strict constraint makes the three-junction design difficult to be realized in experiments\cite{fluxrmp,nori}.
The deficiencies of the one-junction and the three-junction flux qubits urge the development of a third type of flux qubits with both low inductance and weak restriction on the junction parameters. This might be accomplished by exploiting exotic Josephson junctions.

Recently, the topological nanowire Josephson junctions have drawn much attention because the zero energy Majorana bound states (MBSs) appear in the ends of the wires\cite{kitaev,kanermp,beenakker,alicea2}.
The MBSs are potentially useful for fault-tolerant quantum computation: two MBSs can form one Majorana qubit, and the braiding of the MBSs can rotate the Majorana qubit\cite{kanermp,beenakker,alicea2,kitaev,ivanov,aliceanphy}.
In Topological junctions, MBSs show several unique features which can be detected in experiments\cite{law09,lutchyn,oreg,beenakker-disorder,lawnc}.
One important feature of MBSs is that they
transport $4\pi$ period supercurrents, known as the $4\pi$ period fractional Josephson effect\cite{kitaev,lutchyn,aguado}. This novel Josephson effect is regarded as a compelling evidence for the existence of MBSs\cite{alicea2,lutchyn,aguado}. Besides, it can be used to couple the Majorana qubits and the superconducting qubits coherently\cite{pekker}. This coherent coupling can construct quantum gates for Majorana qubits, which is important for the quantum computation\cite{hassler,hassler2,jiang,bonderson,pekker}.
The $4\pi$ period Josephson effect has been experimentally reported in topological nanowire Josephson junctions, where indirect ac signals were observed\cite{williams,rokhinson}.
The direct experimental observation of the fractional Josephson effect
is hindered by several factors. The first one is the existence of a competing $2\pi$ period supercurrent from other Andreev bound states\cite{beenakker91,beenakker11,law11}, which might be even larger than the $4\pi$ period component.
In addition, the small coupling between distant MBSs will induce a parity flipping in the topological junction, which diminishes the $4\pi$ period signal in the dc current\cite{kitaev}.

In this work, we propose a two-junction flux qubit using the $4\pi$ period Josephson effect. For this purpose, we adopt a setup as sketched in Fig. 1. A conventional superconductor-insulator-superconductor Josephson junction is connected with a topological nanowire Josephson junction, forming a small SQUID with vanishing self-inductance.
In the middle of the nanowire junction, a tunneling barrier is created by an external voltage gate.
Four MBSs exist in the wire: $\gamma_2$ and $\gamma_3$ locate near the tunneling barrier, $\gamma_1$ and $\gamma_4$ locate at the ends of the wire.
The tunneling between $\gamma_1$ and $\gamma_2$ contributes a $4\pi$ period component to the Josephson energy of nanowire junction.
This $4\pi$ period Josephson effect induces a double-well potential in the system, under which two lowest energy levels are isolated from other high energy levels.
These two isolated levels present contradicting circulating supercurrents, thus making up a flux qubit. We
study the the system with both analytical and numerical methods, and find that the flux qubit exists in a wide junction parameter range.
Therefore, our proposal provides a third type of flux qubit with low inductance and large variability in junction parameters, which bring in
the advantage of resistance to the external noise and robustness to system parameter modulations.
Moreover, we show that this flux qubit can be coupled with the Majorana qubit when the junction parameters are well tuned. This coupling is useful for manipulating the Majorana qubit, which is important for the topological quantum computation.

This work is organized as follows.
We present the model in section II, and simplify the the system to an equivalent one particle Hamiltonian.
In section III we study the potential energy of the system, and shows that double-well structure exists.
In section IV we discuss the quantum tunneling within the double-well potential and analyze the parameter regime for a flux qubit.
In section V, we show the numerical simulations on the eigenstates of the system, and discuss the energy structures of the flux qubit.
In section VI, we discuss the coupling between the flux qubit and the Majorana qubit. Finally we give discussions and a conclusion in section VII and VIII.

\section{Model}
The setup shown in Fig. 1 is built by a conventional insulator Josephson junction and a topological Josephson junction.
The topological junction is constructed with a spin-orbit coupling nanowire between two conventional s-wave superconductors. The nanowire enters the topological superconducting phase with
appropriate Zeeman energy and chemical potential. In the middle of the wire,
a voltage gate is applied to produce a tunneling barrier.
Four MBSs exist in the wire, two of them locate at the ends of the wire and the other two locate near the tunneling barrier.
The low energy Hamiltonian of the nanowire junction is determined by the the superconducting phase difference across the junction and the quantum states of the MBSs\cite{hassler,jiang,bonderson,pekker},
\begin{eqnarray}
\mathcal{H}_{m} = && - E_{c1}\partial^2_{\theta_m} - J_m\cos\theta_m + i E_m\cos\frac{\theta_m}{2}\gamma_2\gamma_3 \nonumber\\
&& + i\delta_{\rm L}\gamma_1\gamma_2 + i\delta_{\rm R}\gamma_3\gamma_4,
\end{eqnarray}
where $\gamma_2$ and $\gamma_3$ are the two MBSs near the tunneling barrier, $\gamma_1$ and $\gamma_4$ are the MBSs at the ends of the wire, $\theta_m$ is the phase difference, $E_{c1} = e^2/2C_1$ is the charging energy from the capacitance $C_1$ of the nanowire junction, $\delta_{L}$ and $\delta_R$ represents the small wave-function overlaps between the two MBSs at the left and the right sides of the tunneling barrier, respectively.
$J_m$ and $E_m$ account for the $2\pi$ and $4\pi$ period Josephson energy from the quasi-particle channels and the MBS channel.
The $J_m$ might be larger than $E_m$, since we have large number of quasi-particle channels and only one MBS channel.
This produces one of the obstructions for the successful detection of the $4\pi$ period Josephson effect.
We note that the $4\pi$ period Josephson energy will induce an $2\pi$ period dc Josephson current after adding the MBSs coupling $\delta_{L,R}$. In this sense, a $4\pi$ period dc supercurrent does not exist. The $4\pi$ period Josephson effect actually refers to the Josephson energy shown in Eq. (1).

The conventional insulator Josephson junction has been extensively investigated, with a low energy Hamiltonian given as\cite{fluxrmp},
\begin{eqnarray}
\mathcal{H}_{J} = - E_{c2}\partial^2_{\theta_J} - J\cos\theta_J,
\end{eqnarray}
where $\theta_J$ is the phase difference, $E_{c2} = e^2/2C_2$ is the charging energy, and $J$ is the Josephson energy.
A magnetic flux is applied through the SQUID, which induces a inductive energy\cite{fluxrmp},
\begin{eqnarray}
\mathcal{H}_{L} = \frac{\Phi^2_0}{8\pi^2L}(\theta_J - \theta_m - 2\pi\Phi/\Phi_0)^2,
\end{eqnarray}
where $L$ is the self-inductance of the SQUID, $\Phi$ is the applied flux through the loop measured in the unit of magnetic flux quanta $\Phi_0 = h/2e$.
In our setup, we hope to minimize the environmental magnetic noises. Therefore, we adopt a small SQUID with a vanishing self-inductance $L \approx 0$.
 In this scenario, the inductive energy should always be minimized by a constraint\cite{fluxrmp} between $\theta_m$ and $\theta_J$,
\begin{eqnarray}
\theta \equiv \theta_m = \theta_J - 2 \pi \Phi/\Phi_0.
\end{eqnarray}
Under this constraint, only one phase difference $\theta$ is enough to describe the system, drastically simplifying the problem.
Adding together, we arrive at a total Hamiltonian for the system,
\begin{eqnarray}\label{totalH}
\mathcal{H} = && - E_{c}\partial^2_{\theta} - J \cos (\theta + 2\pi\Phi/\Phi_0) - J_m\cos{\theta} \nonumber\\
&& + i E_m\cos\frac{\theta}{2}\gamma_2\gamma_3 + i\delta_{\rm L}\gamma_1\gamma_2 + i\delta_{\rm R}\gamma_3\gamma_4,
\end{eqnarray}
where $E_c = E_{c1} + E_{c2}$ is the total charging energy.

The four MBSs in the nanowire form two Majorana qubits which reflect the parity of the superconducting ground state of the nanowire. The topological parity states is clearly presented by defining complex fermionic operators with the Majorana operators,
\begin{eqnarray}
f^\dagger_1 = (\gamma_4 - i\gamma_1)/2, f^\dagger_2 = (\gamma_2 - i\gamma_3)/2.
\end{eqnarray}
These two fermionic operators are equivalent to the four MBS. Their eigenstates $|0\rangle$, $f^\dagger_1|0\rangle$, $f^\dagger_2|0\rangle$, and $f^\dagger_1f^\dagger_2|0\rangle$  correspond to the superconducting ground states with different fermionic parities.
The total parity of the system should be conserved at low temperature, thus we can decompose the Hilbert space of the Hamiltonian (\ref{totalH}) into two disconnected sub-spaces with even or odd total parity.
 We take the even total parity without losing generality, then the subspace is expanded by $|0\rangle_1 |0\rangle_2$ and $f^\dagger_1f^\dagger_2|0\rangle$.
It is instructive to reformulate the parity states into pseudo-spin states $\mid \downarrow\rangle = |0\rangle_1 |0\rangle_2 $ and $\mid \uparrow\rangle =f^\dagger_1f^\dagger_2|0\rangle$.
In this representation, the Majorana operators can be transformed into pseudo-spin operators $i\gamma_2 \gamma_3 = \sigma_z$ and $i\gamma_1 \gamma_2 = -i\gamma_3 \gamma_4=  \sigma_x$, and the Majorana qubits are transformed into spin qubits. Now the Hamiltonian is rewritten as,
\begin{eqnarray}
\mathcal{H}
 = && - E_{c}\partial^2_{\theta} - J\cos{(\theta + 2\pi\Phi/\Phi_0)} - J_m\cos{\theta} \nonumber\\
&& + \sigma_z E_m\cos\frac{\theta}{2} + \delta_m\sigma_x,
\end{eqnarray}
where $\delta_m = \delta_L - \delta_R$.
The Hamiltonian is now equivalent to a spin one half particle using $\theta$ as the coordinate. The particle has a mass of $m = \hbar^2/2E_c$, moving under electro-static potential $-J\cos{(\theta + 2\pi\Phi/\Phi_0)} - J_m\cos{\theta}$ and a magneto-static potential $\sigma_z E_m\cos\frac{\theta}{2} + \delta_m\sigma_x $.
We can write down the Hamiltonian explicitly in a $2\times2$ form using the spin states in z direction as the basis,
\begin{eqnarray}\label{ph}
\mathcal{H} = \left( \begin{array}{cc}
\frac{P^2_{\theta}}{2m} + U^+(\theta) & \delta_m \\
\delta_m & \frac{P^2_{\theta}}{2m} + U^- (\theta)
\end{array}\right),
\end{eqnarray}
where $P_\theta = i\hbar \partial_\theta$ is the momentum operator for $\theta$, $U^\pm (\theta) = - J\cos(\theta + 2\pi\Phi/\Phi_0) - J_m\cos{\theta}\pm E_m\cos\frac{\theta}{2}$ corresponds to the potentials for the up- and down-spins respectively, where $U^+ (\theta) = U^-(\theta + 2\pi)$. The off-diagonal term $\delta_m$ accounts for the spin flipping strength from the wave-function overlap between MBSs, which is small compared with the Josephson energy $J$ for a long nanowire.
The same effective Hamiltonian can be obtained for odd total parity sub-space by simply taking $\delta_m = \delta_L + \delta_R$.

\begin{figure*}[t]
\begin{center}
\includegraphics[clip = true, width = \columnwidth]{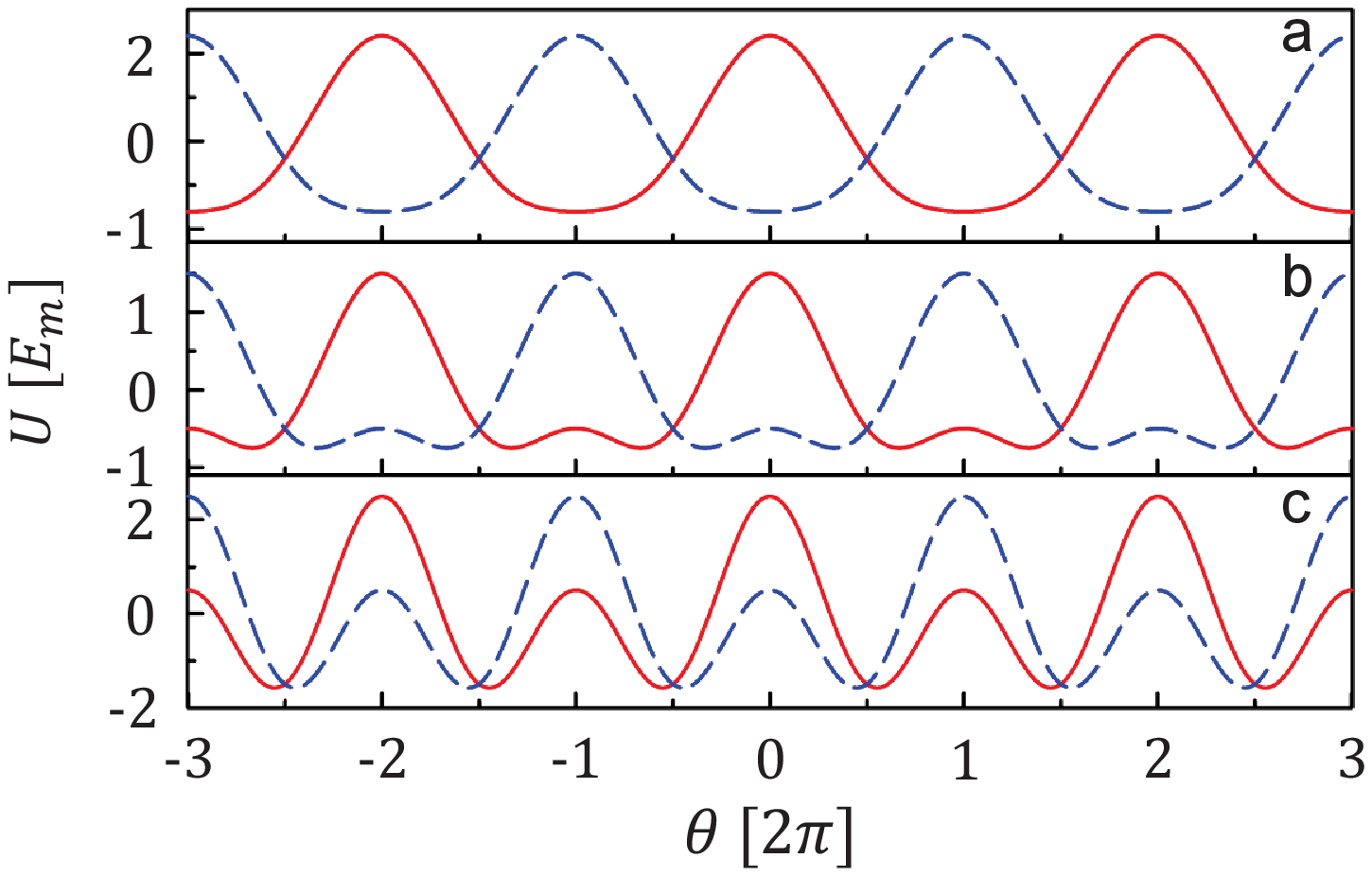}
\includegraphics[clip = true, width = \columnwidth]{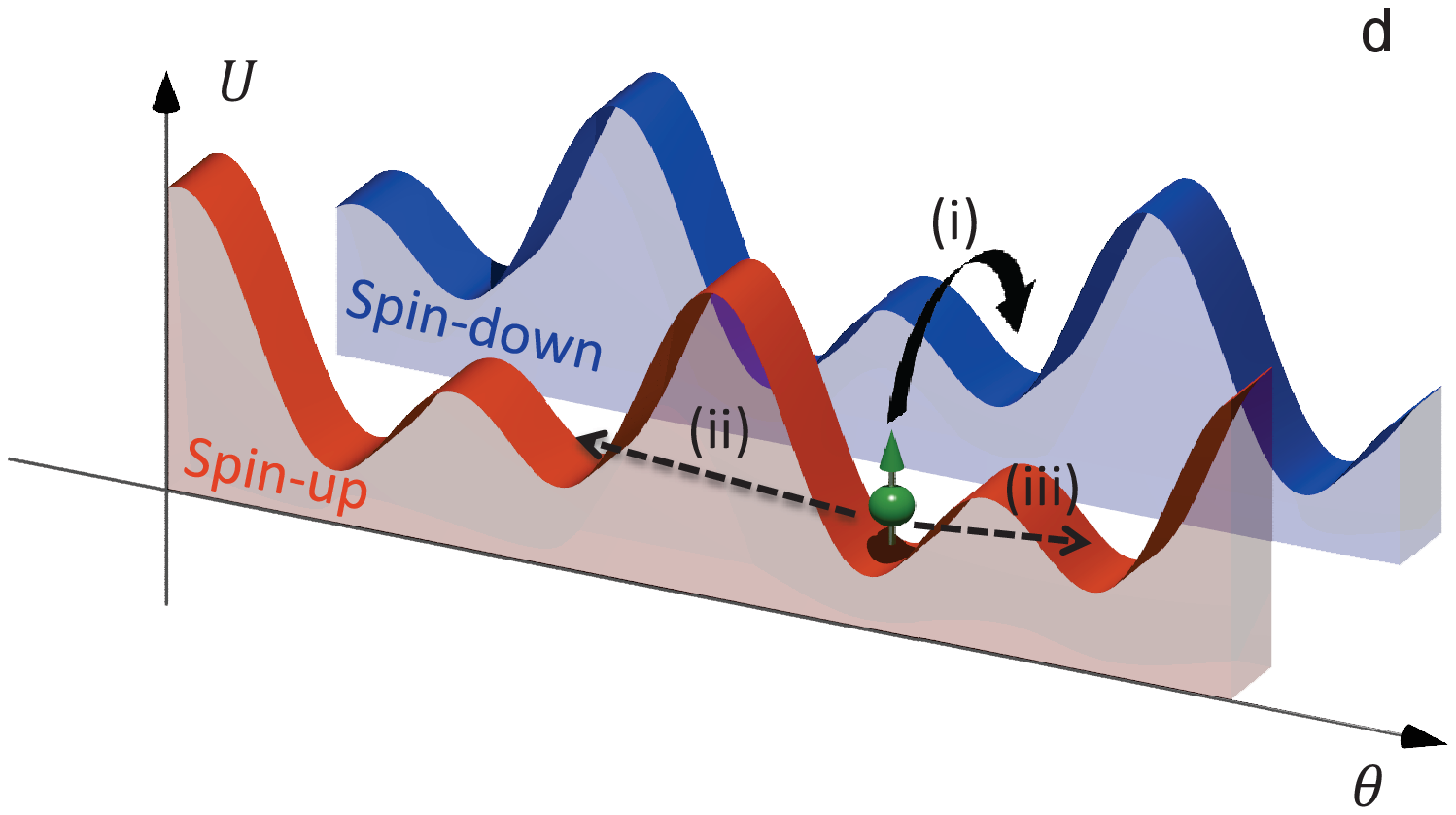}
\caption{(Color online). Potential functions $U^+(\theta)$ (solid line) and $U^-(\theta)$ (dashed line) as a function of superconductor phase difference $\theta$, with (a) $\alpha = 0.2$, (b) $\alpha = 0.5$, and (c) $\alpha = 1.5$. (d) Schematic illustration of three nearest neighbour quantum tunneling process: (i) spin-flipping, (ii) out of the double-well, and (iii) within the double-well.  }
\end{center}
\end{figure*}

\section{Double well potential}
As pointed out in previous studies\cite{fluxrmp,nori,orlando}, a double-well potential is the key ingredient for isolating two energy levels in the one particle Hamiltonian established in Eq. (8).
To obtain a double-well potential, we consider half a quanta of applied magnetic flux through the SQUID. $\Phi = \Phi_0/2$, then the potential $U^\pm (\theta)$ can be written as,
\begin{eqnarray}
U^\pm(\theta) = E_m(\alpha\cos{\theta}\pm\cos{\frac{\theta}{2}}),
\end{eqnarray}
where $\alpha = (J - J_m)/E_m$.
We find that the 2$\pi$ period component $J_m$ in the topological junction is canceled by $J$ in the conventional junction, through destructive interference induced by the magnetic flux $\Phi$.
When $J$ is tuned close to $J_m$, the $4\pi$ period component $E_m$ will make significant contribution to the total Josephson energy, even though it is smaller than $J_m$.
We show the potential $U^\pm(\theta)$ for $\alpha = 0.2$, $0.5$, and $1.5$ in Fig. 2a, 2b and 2c, respectively. The potential for small $\alpha$ has no double-well structure as shown in Fig. 2a . In Fig. 2b, we immediately find the double-well potential structure, which is required for a flux qubit. Figure 2c indicate that the double-well potential still exists when $\alpha$ becomes larger, however, the potential barrier within one double-well increases and approaches the tunneling barrier between different double-wells. We hope to
clarify the parameter regime for the double-well potential structure. Therefore, we study the classical stability points which are given by the zeros in the first derivative of the potentials,
\begin{eqnarray}\label{potential}
\frac{\partial U^\pm(\theta)}{\partial\theta} = - 2E_m\sin\frac{\theta}{2}\left(\alpha\cos{\frac{\theta}{2}}\pm \frac{1}{4}\right) = 0.
\end{eqnarray}
We immediately obtain two sets of solutions for this equation when $\alpha>0$: $\theta = 2n\pi$ with an integer $n$; and $\theta = \pm 2\arccos{(\pm 1/4\alpha)} + 4n\pi$ only for $\alpha > 1/4$.
Now the parameter space is divided into two regimes by a critical value $\alpha_c = 1/4$.
In the regime of $\alpha > \alpha_c$, we have two minimum points and two maximum points, which is the double-well structure as shown in Fig. 2b and 2c.
In the regime of $\alpha < \alpha_c$, there are only one minimum point and one maximum point as shown in Fig. 2a, then the double-well structure disappears. For $\alpha < 0$, we would have the potential up-side-down, and double-well structure does not exist.

In our proposal, we exploit the destructive interference in the SQUID to reduce the $2\pi$ period component in the Josephson energy, and highlight the $4\pi$ period component.
We note that the Josephson energy of the conventional junction plays a two-fold role in this setup.
First, it cancels the $2\pi$ period components in the Majorana junction.
Second, it interplays with the $4\pi$ period part to create the desired double-well potential energy landscape.

\section{Quantum Tunneling}

We want to construct a flux qubit, thus take the regime of $\alpha > \alpha_c$ where the double-well structure exists in the potential $U^\pm (\theta)$.
 We consider the local states at the minimums of the potential, and analyze the energy spectrum due to the quantum tunneling between adjacent local minimums, within the standard tight binding formalism\cite{orlando}. In usual double-well potential systems,
 the two bonding and anti-bonding states should be isolated in the double well, where
the probability of the quantum tunneling out of a double-well is negligible.
In this system, however, one subtlety must be addressed: the particle has a spin, while the spin-up and the spin-down particle have different potentials. As a result, we have to consider the quantum tunneling from the spin-flipping process induced by $\delta_m$.
In the following, we study all three types of nearest neighbor quantum tunneling as sketched in Fig. 2d: (i) the spin-flipping quantum tunneling between the nearby local potential minimums for different spins, (ii) the tunneling out of a double-well in one spin branch, and (iii) the tunneling within one double-well.
We will calculate the three hybridization energies from each quantum tunneling processes, and find the parameter regimes in which two of them can be neglected.

We start from the spin-flipping quantum tunneling between the nearby local states of different spin branches.
The hybridization energy is determined by the overlap of the two local wave functions, multiplied by the spin-flipping energy $2\delta_m$,
\begin{eqnarray}
E_1 \approx 2\delta_m\chi(\theta_1,\theta_2) = 2\delta_m\int d\theta\psi^*(\theta,\theta_1)\psi(\theta,\theta_2),
\end{eqnarray}
where $\theta_1 = 2\arccos{(1/4\alpha)}$ is a local minimum for $U^-(\theta)$, and $\theta_2 = 2\pi - \theta_1$ is the local minimum for $U^+(\theta)$ close to $\theta_1$; $\psi(\theta,\theta_{1,2})$ are the lowest energy local wave-functions at the potential minimum $\theta_{1,2}$, which can be calculated with a harmonic oscillator approximation.
We define the harmonic potential $V(\theta) = \frac{1}{2} K(\theta - \theta_{1,2})^2 + V_0$, where $V_0 = - E_m(\alpha + 1/8\alpha)$ corresponds to the minimum of $U^\pm (\theta)$, and $K$ is given by,
\begin{eqnarray}
K = \frac{\partial^2 U^- (\theta)}{\partial \theta^2}|_{\theta = \theta_1} = E_m(\alpha - \frac{1}{16\alpha}).
\end{eqnarray}
The local wave function is obtained by solving the Shr\"{o}dinger equations under the harmonic potential,
\begin{eqnarray}
\psi(\theta,\theta_{i}) = (\frac{K}{2\pi^2E_c})^\frac{1}{8} e^{-\sqrt{\frac{K}{8E_c}}(\theta - \theta_{i})^2},
\end{eqnarray}
with $i=1,2$. Then the wave-function overlap is given by,
\begin{eqnarray}
\chi(\theta_1,\theta_2) = e^{-\sqrt{\frac{K}{32E_C}}(\theta_1 - \theta_2)^2}.
\end{eqnarray}
Plugging this wave function overlap back to Eq. (11), we can obtain the hybridization energy from the spin-flipping tunneling.

\begin{figure}[t]
\begin{center}
\includegraphics[clip = true, width = \columnwidth]{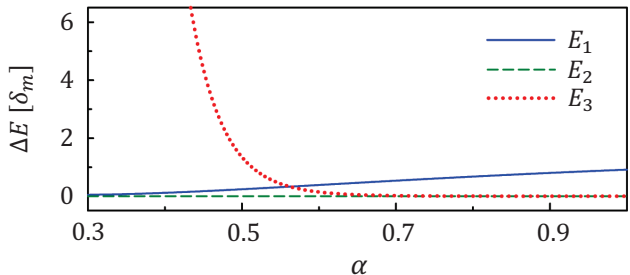}
\caption{(Color online). Hybridization energies $E_{1}$ (solid line), $E_{2}$ (dashed line), and $E_{3}$ (dotted line) as a function of $\alpha$, with $E_c = 0.05 E_m$ and $\delta_m = 0.0001E_m$.}
\end{center}
\end{figure}

Second, we consider the hybridization energy from the quantum tunneling out of a double-well potential.
We use the WKB method to solve the problem, which was widely used for treating the quantum tunneling in Josephson junction systems\cite{orlando}. The resulted hybridization energy is
\begin{eqnarray}
E_2 \approx \frac{\omega\hbar}{\pi}\exp \left(- \frac{1}{\sqrt{E_c}}\int_{\theta_1}^{4\pi - \theta_1} d\theta\sqrt{U^-(\theta) - V_0}\right),\nonumber\\
\end{eqnarray}
where $\omega = \sqrt{2K E_c}/\hbar$ is the frequency of ground state within the harmonic oscillator approximation. The integration on the exponential function represents the action of the particle moving from one minimum potential point to another.
With the same formalism, we can also obtain the hybridization energy from the quantum tunneling within one double-well,
\begin{eqnarray}
E_3 \approx \frac{\omega\hbar}{\pi}\exp \left(-\frac{1}{\sqrt{E_c}}\int_{-\theta_1}^{\theta_1} d\theta\sqrt{U^-(\theta) - V_0}\right).
\end{eqnarray}
We notice that the formulas for the two hybridization energies $E_2$ and $E_3$ are identical, except different actions from the different integration zones.
The value $E_2/E_3$ is depending on the potential landscape, which is controlled by $\alpha$. We expect that $E_2/E_3$ should approaches one for large $\alpha$ and approaches zero for small $\alpha$.
$E_1$ is proportional to $\delta_m$ which is an exponential function of the nanowire length. Therefore, $E_1$ will also be very different for long and short nanowires.
Two of these three energies should be vanishingly small to construct a flux qubit, {\it i.e.} $E_3 \gg E_{1,2}$ in this system.
The result of $E_{1,2,3}$ is shown in Fig. 3, as a function of $\alpha$ which is the control parameter in this setup. As expected, we find that $E_3$ increases quickly when $\alpha$ decreases.
For $1/4<\alpha < 1/2$, $E_3$ is larger than $E_1$ and $E_2$ by orders, thus we can safely ignore $E_1$ and $E_2$ in this region. We get a one-particle system tunneling within an isolated double-well potential, and should obtain two lowest energy states which are the
bonding and the anti-bonding states separated by $E_3$ in energy.

\begin{figure}[t]
\begin{center}
\includegraphics[clip = true, width = \columnwidth]{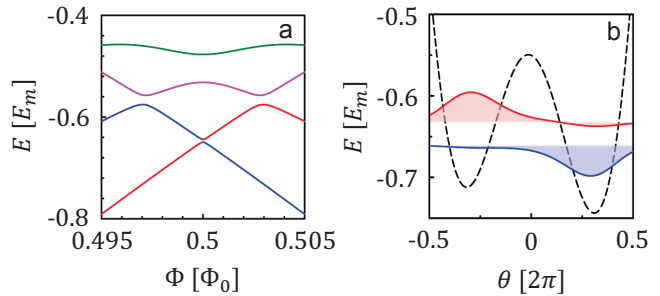}
\caption{(Color online). (a) Four lowest energy levels as a function of $\Phi$, with $\alpha = 0.45$, $E_c = 0.05E_m$, $\delta_m = 0.0001E_m$, and $J_m=5E_m$. (b) Wave functions of the two lowest energy levels with $\Phi = 0.5005\Phi_0$; the dashed line is the potential function $U^-(\theta)$.}
\end{center}
\end{figure}

\section{Flux Qubit}
The analytic results shown in Fig. 3 are obtained for exactly half quanta of applied magnetic flux through the SQUID, where the double-well potentials are symmetric. In this case, the eigenstates are bonding and anti-bonding states for which the wave functions distribute equally at the two wells.
If the applied magnetic flux deviate a little bit, the double-well becomes asymmetric.
Then the lowest energy state should stay at the deeper well, while the other state should stay at the other well.
We numerically solve the Hamiltonian in Eq. (\ref{ph}) to obtain the energy spectra. We apply both the finite difference method in real space and the plane wave expansion method in momentum space, and find identical results.
 The four lowest energy levels of the spin-down particle as a function of applied magnetic flux $\Phi$ is shown in Fig. 4a.
It can be seen that two lowest energy levels are isolated from other higher energy levels around $\Phi = \Phi_0/2$.
The two levels show opposite slopes as a function of $\Phi$, indicating that the supercurrents carried by them have opposite circulating directions.
To confirmation this, we calculate the wave functions of these two levels at $\Phi= 0.5005\Phi_0$, as shown in Fig. 4b.
As expected, we find that the two states stay at two different $\theta$s, which account for contradicting supercurrents across the junctions.
These numerical results demonstrate that the SQUID works effectively as a two-level system, with opposite supercurrents assigned to each level. We obtain identical energy levels for the spin-down particle. In fact, all levels are two-fold degenerate due to the spin degree of freedom.
After checking the details of the wave functions, we find that the two lowest energy levels around $\Phi = 0.5\Phi_0$ can be described by an effective Hamiltonian,
\begin{eqnarray}
H = [A(\Phi)\tau_z + B\tau_x ]\sigma_z,
\end{eqnarray}
where $B \approx E_3/2$ determines the energy splitting at $\Phi=\Phi_0/2$, and $A(\Phi) \approx  \frac{\partial U^-(\theta_1,\Phi)}{\partial \Phi}|_{\Phi = \Phi_0/2} (\Phi- \Phi_0/2) =  [- \pi J  \sqrt{16\alpha^2 - 1}/4\alpha^2] (\Phi/\Phi_0 - 1/2)$. Here $\tau_{x,z}$ are the pauli matrices with the two supercurrent circulating directions as the basis states. The spin is conserved when the flux $\Phi$ changes in the effective Hamiltonian Eq. (17), because the spin-flipping term is negligible comparing with the dominating energy scale $B=E_3/2$.
We obtain a typical effective Hamiltonian for a flux qubit, where the two basis states are represented by contradicting supercurrents.

We note that these numerical results are not sensitive to our present choice of junction parameters.
In fact, similar results can be obtained for a range of parameters.
From the quantum tunneling analysis, we find that the Josephson energy ratio $\alpha$ should be larger than $1/4$ to keep the double-well potential, but smaller than $1/2$ to achieve the isolation of one double-well.
Our numerical results are consistent with the theoretical analysis, the two isolated levels described by Eq. (17) indeed exist within this range.

\section{Coupling the flux qubit to the Majorana qubit}
Looking back at Fig. 3, we notice that the spin-flipping hybridization energy $E_1$ becomes important for large $\alpha$, since the other two hybridization energy become vanishingly small.
Therefore, isolated low energy states should also exist. In this regime, however, the spin-flipping process plays the key role. Therefore, the spin is not conserved when changing the magnetic flux $\Phi$.
We confirm this picture by numerical simulations, and show the results of the energy spectrum with $\alpha = 1$ in Fig. 5a.
We find two lowest energy levels which become energetically isolated from higher energy levels around $\Phi = \Phi_0/2$.
We further study the wave functions of these two levels at $\Phi = 0.501\Phi_0$, and show the results in Fig. 5b.
These two states exhibit contradicting supercurrent, therefore can be used as a flux qubit. Importantly, they have opposite spins, reflecting a spin hybridization as sketched in Fig. 5a.
We check the details of the wave function of the states, and obtain an effective Hamiltonian for the two lowest energy levels around $\Phi= \Phi_0/2$,
\begin{eqnarray}
H = A(\Phi)\tau_z \sigma_z + B'\tau_x \sigma_x ,
\end{eqnarray}
where $B' = E_1/2$. Unlike the pure flux qubit obtained in previous section, we arrive at a hybrid qubit where the flux qubit is coupled with the Majorana qubit (spin qubit). This hybrid qubit exists in the range of $\alpha>0.7$ for the current choice of MBSs coupling $\delta_m$. If $\delta$ increases or decreases, the range of $\alpha$ should change accordingly.

\begin{figure}[t]
\begin{center}
\includegraphics[clip = true, width = \columnwidth]{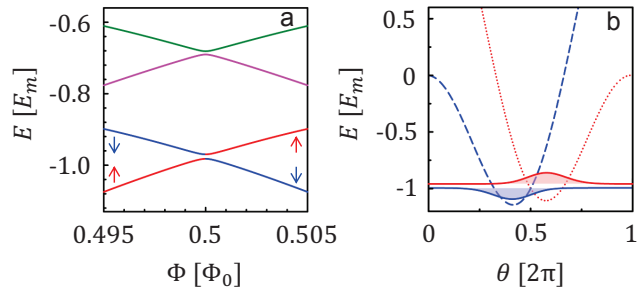}
\caption{(Color online). (a) Four lowest energy levels as a function of $\Phi$, with $\alpha = 1$, $E_c = 0.05E_m$, $\delta_m = 0.015E_m$, and $J_m=5E_m$. The arrows show the spins of the two lowest energy levels. (b) Wave functions of the two lowest energy levels with $\Phi = 0.501\Phi_0$; the dashed line and the dotted line are the potential functions $U^-(\theta)$ and $U^+(\theta)$, respectively.}
\end{center}
\end{figure}

\section{Discussions}

Our proposal is based on a topological nanowire Josephson junction. However, the idea could be naturally extended to other topological junctions where the $4\pi$ period Josephson effect exists. In particular, the quantum spin hall insulator in proximity to superconductivity is a promising candidate\cite{beenakker13}. In addition, the $4\pi$ Josephson period effect was also argued to be existing in topologically trivial Josephson junctions\cite{golubov}, such as ballistic point contact junction. Our proposal should also work for these Josephson junctions.

The proposed two-junction flux qubit could be realized with present experimental technologies. The key ingredient of the proposal, a topological superconducting nanowire Josephson junction, has been reported in recent experiments\cite{williams,rokhinson,kouwenhoven}. The growth of a conventional Josephson junction with controllable critical current was realized in several systems\cite{golubov}. The superconducting circuit which connects these two junctions into a SQUID could be achieved by the lithography technology, with controllable inductance and capacitance\cite{clarke}. Moreover, these devices can be integrated easily\cite{fluxrmp,dwave}, therefore, the proposed flux qubit can be scaled to a large number.
The typical value of the junction parameters are
$E_m \sim 5 \mu eV$, $E_c \sim 0.1 \mu eV$. The size of the SQUID should be in the range of micrometers with a vanishing self-conductance of $L \sim 1 p{\rm H}$. The coupling energy of the MBSs depends on the length of the wire,
with $ \delta \sim 1 neV$ for long wire and $\delta \sim 0.1 \mu eV$ for short wire. The operation temperature should be small to achieve quantum coherence, with an estimation of $T \sim 50 mK $.

This two-junction flux qubit can be measured by an external quantum circuit which detects the magnetic flux in the qubit accurately and sensitively\cite{fluxrmp,nori}. Meanwhile,
it can be manipulated by applying an electro-magnetic pulse. When the frequency of the pulse fits the energy splitting between the two eigenstates of the flux qubit, the Rabi oscillation will drive a coherent rotation on the qubit state\cite{fluxrmp}. These operations on the two-junction flux qubits do not require severer conditions than other types of flux qubits, thus should be experimentally realistic.

We note that the operations on the flux qubit are based on electro-magnetic interactions. In this case, the Majorana qubit should be untouched during the operation since MBSs are charge neutral quasi-particles. This is correct for the pure flux qubit regime shown in Eq. (17). In this regime, the Majorana qubit is not coupled with the flux qubit, thus should remain the same under electro-magnetic perturbations. However, the situation is different for the hybrid qubit regime shown in Eq. (18). The coupling between the flux qubit and the Majorana qubit induces a strong correlation between flux state and the spin state, then the Majorana qubit can be manipulated by a rotation of the flux qubit. This manipulation method for the Majorana qubit might be useful for the topological quantum computation.

\section{Conclusion}

We propose a two-junction flux qubit using the $4\pi$ period Josephson effect. We reveal that this flux qubit shares all the merits of the previous two types of flux qubits, yet avoids their disadvantages.
The proposed flux qubit is made of a small dc SQUID, formed by a conventional Josephson junction and a topological nanowire Josephson junction.
The SQUID is built with a low inductance to reduce the external noises. In this regime, the system is fully described by a phase difference $\theta$.
The dynamics of the system resembles a spin one half particle moving in one dimension space, with $\theta$ as the particle coordinate. The charging energy from the capacitance acts as the kinetic energy of the particle, and the Josephson energy acts as the potential energy.
We investigate the Josephson energy of the system, and find that the $2\pi$ period component in the nanowire junction can be annihilated by a destructive interference with the conventional junction. Then, the $4\pi$ period component in the nanowire dominates the system, bringing in a double-well shape potential.
We study the low energy eigenstates of the system, and reveal that the double-well potential induces the isolation of two levels.
These two levels have contradicting circulating supercurrents, thus can be used as a flux qubit.
Furthermore, we find that a fine tuning of the SQUID will result in an coupling between the flux qubit and the Majorana qubit in the nanowire junction. This coupling might be important for quantum computation.
In summary, our proposal provides a two-junction flux qubit which has the merits of low inductance and large variability in junction parameters.

\section*{Acknowledgements}

This work was supported by NSFC-11304400, NSFC-61471401, SRFDP-20130171120015, and 985 Project of Sun Yat-Sen University.
D.X.Y. is supported by the National Basic Research Program of China (2012CB821400), NSFC-11074310, NSFC-11275279, SRFDP-20110171110026, and Fundamental Research Funds for the Central Universities of China.


\begin{thebibliography}{00}

\bibitem{Nielsen} M. A. Nielsen, I. L. Chuang, {\it Quantum Computation and Quantum Information} (Cambridege University Press, Cambridge, England, 2000).
\bibitem{DiVincenzo} C. H. Bennett and D. P. DiVincenzo, Nature \textbf{404}, 247 (2000).
\bibitem{Lloyd} S. Lloyd, Science \textbf{273}, 1073 (1996).
\bibitem{Schumacher} B. Schumacher, Phys. Rev. A \textbf{51}, 2738 (1995).
\bibitem{spin} D. P. DiVincenzo, Science \textbf{269}, 225 (1995).
\bibitem{QED} Q. A. Turchette, C. J. Hood, W. Lange, H. Mabuchi, and H. J. Kimble, Phys. Rev. Lett. \textbf{75}, 4710 (1995).
\bibitem{charge} Y. Nakamura, Yu. A. Pashkin, J. S. Tsai, Nature \textbf{398}, 786 (1999).
\bibitem{flux3} J. E. Mooij, T. P. Orlando, L. Levitov, Lin Tian, Caspar H. van der Wal, Seth Lloyd, Science \textbf{285}, 1036 (1999).
\bibitem{flux1} J. R. Friedman, V. Patel, W. Chen, S. K. Tolpygo, and J. E. Lukens, Nature \textbf{406}, 43 (2000).
\bibitem{dwave} S. Boixo, T. F. R{\o}nnow, S. V. Isakov, Z. H. Wang, D. Wecker, D. A. Lidar, J. M. Martinis, and M. Troyer, Nature Physics \textbf{10}, 218 (2014).
\bibitem{fluxrmp} Y. Makhlin, G. Sch\"{o}n, A. Shnirman, Rev. Mod. Phys. \textbf{73}, 357 (2001).
\bibitem{nori} J. Q. You and F. Nori, Physics Today \textbf{58}, 42 (2005).
\bibitem{clarke} J. Clarke and F. K. Wilhelm, Nature \textbf{453}, 1031 (2008).
\bibitem{devoret} M. H. Devoret and R. J. Schoelkopf, Science \textbf{339}, 1169 (2013).
\bibitem{kitaev} A. Kitaev, Phys. Usp. \textbf{44}, 131 (2001).




\bibitem{kanermp} M. Z. Hasan, C. L. Kane, Rev. Mod. Phys. {\bf 82}, 3045 (2010).
\bibitem{beenakker} C. W. J. Beenakker, Annu. Rev. Con. Mat. Phys. \textbf{4}, 113 (2013).
\bibitem{alicea2} J. Alicea, Rep. Prog. Phys. 75, 076501 (2012).

\bibitem{ivanov} D. A. Ivanov, Phys. Rev. Lett. {\bf 86}, 268 (2001).
\bibitem{aliceanphy} J. Alicea, Y. Oreg, G. Refael, F. von Oppen and M. P. A. Fisher, Nature Physics \textbf{7}, 412 (2011).


\bibitem{law09} K. T. Law, P. A. Lee, and T. K. Ng, Phys. Rev. Lett. \textbf{103}, 237001 (2009).
\bibitem{lutchyn} R. M. Lutchyn, J. D. Sau, and S. Das Sarma, Phys. Rev. Lett. {\bf 105}, 077001 (2010).
\bibitem{oreg} Y. Oreg, G. Refael and F. von Oppen, Phys. Rev. Lett. \textbf{105}, 177002 (2010).
\bibitem{beenakker-disorder} A. R. Akhmerov, J. P. Dahlhaus, F. Hassler, M. Wimmer, and C. W. J. Beenakker, Phys. Rev. Lett. \textbf{106}, 057001 (2011).
\bibitem{lawnc} J. J. He, J. S. Wu, T. P. Choy, X. J. Liu, Y. Tanaka, and K. T. Law, Nat. Commun. \textbf{5}, 3232 (2014).





\bibitem{aguado} P. San-Jose, E. Prada, and R. Aguado, Phys. Rev. Lett. \textbf{108}, 257001 (2012).


\bibitem{pekker} D. Pekker, C. Y. Hou, V. E. Manucharyan, and E. Demler, Phys. Rev. Lett. \textbf{111}, 107007 (2013).


\bibitem{hassler} F. Hassler, A.R. Akhmerov, C.-Y. Hou, and C.W.J. Beenakker, New J. Phys. \textbf{12}, 125002 (2010).
\bibitem{hassler2} F. Hassler, A.R. Akhmerov, and C.W.J. Beenakker, New J. Phys. \textbf{13}, 095004 (2011).

\bibitem{jiang} L. Jiang, C. L. Kane, and J. Preskill, Phys. Rev. Lett. \textbf{106}, 130504 (2011).
\bibitem{bonderson} P. Bonderson, and R. M. Lutchyn, Phys. Rev. Lett. \textbf{106}, 130505 (2011).




\bibitem{williams} J. R. Williams, A. J. Bestwick, P. Gallagher, S.S. Hong, Y. Cui, A. S. Bleich, J. G. Analytis, I. R. Fisher, and D. Goldhaber-Gordon, Phys. Rev. Lett. \textbf{109}, 056803 (2012).
\bibitem{rokhinson} L. P. Rokhinson, X. Liu, and J. K. Furdyna, Nature Physics \textbf{8}, 795 (2012).
\bibitem{beenakker91} C. W. J. Beenakker, Phys. Rev. Lett. \textbf{67}, 3836 (1991).
\bibitem{beenakker11} B. van Heck, F. Hassler, A. R. Akhmerov, and C. W. J. Beenakker, Phys. Rev. B \textbf{84}, 180502 (2011).
\bibitem{law11} K. T. Law and P. A. Lee, Phys. Rev. B \textbf{84}, 081304 (2011).
\bibitem{orlando} T. P. Orlando, J. E. Mooij, Lin Tian, Caspar H. van der Wal, L. S. Levitov, Seth Lloyd, and J. J. Mazo, Phys. Rev. B \textbf{60}, 15398 (1999).

\bibitem{beenakker13} C. W. J. Beenakker, D. I. Pikulin, T. Hyart, H. Schomerus, and J. P. Dahlhaus, Phys. Rev. Lett. \textbf{110}, 017003 (2013).
\bibitem{golubov} A. A. Golubov, M. Yu. Kupriyanov, E. Il¡¯ichev, Rev. Mod. Phys. \textbf{76}, 411 (2004).

    \bibitem{kouwenhoven} V. Mourik, K. Zuo, S. M. Frolov, S. R. Plissard, E. P. A. M. Bakkers, L. P. Kouwenhoven, Science {\bf 336}, 1003 (2012).
\end{thebibliography}
\end{document}